\documentstyle[preprint,prd,aps]{revtex}
\begin{document}

\draft
\title{{\bf  Quantum state correction of relic gravitons from quantum gravity}}
\author{{\bf J.L. ROSALES }
\footnote{E-mail: rosales@phyq1.physik.uni-freiburg.de}\\
       {\em Fakult\"at f\"ur Physik,\\
	 Universit\"at Freiburg,Hermann-Herder-Strasse 3 \\
	D-79104 Freiburg, Germany.}}
\date{\today}
\begin{center}
\maketitle
\begin{abstract}

	The semiclassical approach  to quantum gravity
	would yield
	the Schr\"odinger formalism for the wave function of metric
	perturbations or gravitons plus quantum gravity correcting terms in
	pure gravity; thus, in the inflationary scenario,
	we should expect correcting effects to the relic
	graviton (Zel'dovich) spectrum of the order of $(\lambda/m_{Pl}^{2})$.
	These, on the other hand, could possibly be measured in a future
	experiment.

\end{abstract}

\end{center}

\section{Introduction}
  In spite of the fact that there still lacks a full {\em quantum gravity}
 theory, a predictive semiclassical {\em quantum cosmology}
 has already been developed due to the selection of  the
 initial quantum state of the Universe. The latter may be developed upon
 quantising canonical general relativity for  the metric tensor 
 corresponding to perturbations about FRW models \cite{kn:Hawking} -
 \cite{kn:Wada}. In such an approximation scheme, the
 parameter representing the relevant energy for the initial expansion
 of the Universe, the effective cosmological constant, $\lambda$, 
 is much lower than the Planck mass \cite{kn:Rubakov},

\begin{equation}
\lambda/m_{Pl}^{2}\leq 10^{-9}.
\end{equation}
Thus, upon using a Born-Oppenhiemer
approximation scheme (that takes into account the above mentioned
relative small value of matter energy with respect to the Planck mass), one
can obtain, for a given complex solution of Wheeler-DeWitt equation,
the Schr\"odinger formalism for the wave function of metric
perturbations \cite{kn:Banks}.
Moreover, {\em quantum gravity} corrections also arise \cite{kn:Kiefer}
and, at least in principle, they should show how effects of quantum gravity
might even be measured in the context of these inflationary scenarios.

The aim of this paper is precisely to show that such effects do exist,
in this case, upon introducing a shift of quantum gravity origin
in the spectrum of the relic gravitons being originated during the early inflationary
stages of the Universe.
Relic gravitons  (Zel'dovich) spectrum has, as a matter of fact,
already been measured from the existing $3K$ cosmic background radiation
anisotropy \cite{kn:Steinhardt}; on the other hand,
a certain quantum gravity modification could be possibly measured
in the next future by some experimental devices \cite{kn:COBRAS}.

\section{Semiclassical wave function in de Sitter scenario}

The simplest (although still realistic)  model for the early stage of
the Universe consists of
a massive scalar field with matter potential $U(\phi)$ in a FRW
spacetime described by a single degree of freedom, the scale factor,
$a(t)$.
After the Hartle-Hawking prescription \cite{kn:Hartle}, the wave function
will depend on the value $U(\phi(\Sigma))$ ( $\Sigma$ being the
boundary of the compact euclidean 4 - sphere) of the inflaton (scalar field)
potential.  This is just de Sitter
case with $U(\phi)\approx \lambda$
playing the role of an effective cosmological constant for
large initial values of $\phi$; thus, in this model,
we only  take into account, for the wave function, an implicit dependence on the
inflaton  field (in classical terms:
we will ignore the backreaction of matter
corresponding to the kinetic energy of the field,
so that matter couples directly to curvature using the effective
cosmological constant).

Moreover, the general approach is only physically consistent if we obtain
a description for the Schr\"odinger evolution of metric perturbations
or gravitons,
i.e., the tensor harmonics of the three sphere $d_{n}$.
Now, upon defining $\alpha=\log (a)$ as the gravitational variable, the
metric perturbation Hamiltonian is
\begin{equation}
H_{m}=\sum_{n}\frac{-1}{2}\frac{\partial^{2}}{\partial d_{n}^{2}}+ (n^{2}-1)e^{4\alpha} d_{n}^{2} \mbox{,}
\end{equation}
where, as a consequence of the perturbative character  of the $\{d_{n}\}$,
\begin{equation}
d_{n}^{2}/\lambda\ll 1
\end{equation}
Therefore, upon assuming $U(\phi)\approx \lambda$,
and neglecting second derivatives with respect to the matter field,
Wheeler-DeWitt equation reads \cite{kn:Wada}
\begin{equation}
[\frac{1}{2m_{Pl}^{2}}\frac{\partial^{2}}{\partial \alpha^{2}}+
\frac{m_{Pl}^{2}}{2} (-e^{4\alpha}+\lambda e^{6\alpha})+H_{m}]\Psi=0
\end{equation}

In the semiclassical approach we  write the wave functional $\Psi[\alpha,\{d_{n}\}]$ as
\begin{equation}
\Psi=e^{iS}
\end{equation}
then we expand $S$ in the form
\begin{equation}
S=m_{Pl}^{2}S_{0}+S_{1}+m_{Pl}^{-2}S_{2}+... \mbox{.}
\end{equation}
   
We now insert the expansion defined by Eqs. (5) and (6) in  Eq.(4) 
and compare
expressions with the same order in $m_{Pl}$. The highest order yields to
\begin{equation}
\sum_{n}(\frac{\partial S_{0}}{\partial d_{n}})^{2}=0
\end{equation}
Thus, $S_{0}$ depends only on $\alpha$ (the three-metric).

The next order leads to the Hamilton-Jacobi equation for gravity alone:
\begin{equation}
-(\frac{\partial S_{0}}{\partial \alpha})^{2}
 +\lambda e^{6\alpha}-e^{4\alpha}=0 \mbox{.}
\end{equation}
If we define a functional $\psi_{0}$ according to ($S_{1}\equiv \sum_{n} S_{1n}$)
\begin{equation}
\psi_{0}\equiv D(\alpha)e^{i(\sum_{n} S_{1n})}= \Pi_{n}\psi_{0n}
\end{equation}
where,
\begin{equation}
D(\alpha)^{2}=(\frac{\partial S_{0}}{\partial \alpha})\equiv S_{0}'
\end{equation}
then order $m_{Pl}^{0}$ will imply that $\psi_{0}$ is the solution of

\begin{equation}
-i\frac{\partial S_{0}}{\partial \alpha}\frac{\partial \psi_{0}}{\partial \alpha}\equiv
i \frac{\partial \psi_{0}}{\partial t}= H_{m}\psi_{0} \mbox{.}
\end{equation}
Which is the functional Schr\"odinger equation for matter fields propagating
on a fixed curved background; $t$ is usually called WKB time.

If we do not take into account the value of $S_{2}$ in the formal 
expansion, we get to this order of approximation the WKB wave function given by

\begin{equation}
\Psi^{(1)}=\frac{1}{D}e^{im_{Pl}^{2}S_{0}}\psi_{0} \mbox{.}
\end{equation}

\section{ Quantum gravity corrections}

Upon defining $S_{2}=\sigma(\alpha)+\eta(\{d_{n}\},\alpha)$,
we obtain the second order correction for the wave function
\begin{equation}
\Psi^{(2)}=\frac{1}{D}e^{im_{Pl}^{2}S_{0}+i\sigma/m_{Pl}^2}\psi_{0}e^{i\eta/m_{Pl}^{2}}
\end{equation}
where $\sigma$ and $\eta$, representing quantum gravity correcting 
terms to the WKB wave function, satisfy, after some algebra\cite{kn:Kiefer}

\begin{equation}
-S_{0}'\eta'=\frac{i}{2}\sum_{n}\frac{\partial ^{2}\eta}{\partial d_{n}^{2}}+
\frac{i}{\psi_{0}}\sum_{n}\frac{\partial\eta}{\partial d_{n}}
\frac{\partial \psi_{0}}{\partial d_{n}}+
\frac{1}{2 S_{0}'\psi_{0}}\psi_{0}'S_{0}''-\frac{1}{2\psi_{0}}\psi_{0}''
\end{equation}

\begin{equation}
\sigma'=-\frac{S_{0}'''}{4S_{0}'^{2}}+\frac{3}{8}\frac{S_{0}''^{2}}{S_{0}'^{3}}
\end{equation}

Eq. (14) involves entanglement of the modes and, therefore,
it is difficult to solve. Moreover, in general,
field modes and gravitational degrees of
freedom are also entangled so we do not expect to obtain its general solution.
Nonetheless, let us define $\eta(\{d_{n}\},\alpha)\equiv
\sum_{n}\eta_{n}(d_{n},\alpha)$. In that case, we obtain, 
\begin{eqnarray*}
-S_{0}'\sum_{n}\eta_{n}'=\sum_{n}\{\frac{i}{2}\frac{\partial^{2}\eta_{n}}{\partial d_{n}^{2}}+
\frac{i}{\psi_{0n}}\frac{\partial\eta_{n}}{\partial d_{n}}
\frac{\partial \psi_{0n}}{\partial d_{n}}+
\frac{1}{2 S_{0}'\psi_{0n}}\psi_{0n}'S_{0}''-\frac{1}{2\psi_{0n}}\psi_{0n}''\}+
\end{eqnarray*}
\begin{equation}
+\frac{i}{2}\sum_{l\neq k}\frac{1}{\psi_{0k}\psi_{0l}}\psi_{0k}'\psi_{0l}'
\end{equation}
where $\psi_{0n}$ is the solution of the Schr\"odinger equation.
The non boundary proposal \cite{kn:Halliwell} \cite{kn:Wada}
 picks up the ground state (these results are desired in the semiclassical
 approach to gravity \cite{kn:Note}),
\begin{equation}
\psi_{0n}(a,d_{n})=N_{n}(a)e^{-\frac{1}{2}n a^{2}d_{n}^{2}}
\end{equation}
and as a result of this, last term in Eq. (16) is $O(a^{4}d_{k}^{2}d_{l}^{2})$ which, after
Eq. (3) should be neglegible for $a^{2}\rightarrow 1/\lambda$. Thus the modes
are exactly disentangled after the selection of this particular
initial condition and we can finally write, for the correcting phase of
a single mode vave function,

\begin{equation}
-S_{0}'\eta_{n}'= \frac{i}{2}\frac{\partial^{2}\eta_{n}}{\partial d_{n}^{2}}+
\frac{i}{\psi_{0n}}\frac{\partial\eta_{n}}{\partial d_{n}}
\frac{\partial \psi_{0n}}{\partial d_{n}}+
\frac{1}{2 S_{0}'\psi_{0n}}\psi_{0n}'S_{0}''-\frac{1}{2\psi_{0n}}\psi_{0n}''
\end{equation}

We must now solve the system of Eqs. (15) and (18) in order to obtain
predictions from the corrected $\Psi^{(2)}$ wave fuction. To this aim, it is better
to consider the very early stages of the Universe where such
terms should be relevant, i.e., we must restrict our calculation to the region
where $\lambda$ is really a constant; now, for $a^{2}\rightarrow 1/\lambda$,
we can opperate the expressions a little bit further.
First, since the theory should not depend on the selection of the time parameter,
we are allowed to make our predictions for the particular semiclassical evolution
parameter given by
\begin{equation}
\theta(a)=\lambda D(a)=\lambda a^{2}(\lambda a^{2} -1)^{1/2}
\end{equation}
thus, for $a^{2}=1/\lambda (1+\theta^{2}+ O(\theta^{4}))$, we get, after Eqs.
(18) and (15)
\begin{equation}
\frac{i}{\lambda}\frac{\partial \eta_{n}}{\partial \theta}\approx
-\frac{1}{2}\frac{\partial^{2}\eta_{n}}{\partial d_{n}^{2}}-
\frac{1}{\psi_{0n}}\frac{\partial \eta_{n}}{\partial d_{n}}
\frac{\partial \psi_{0n}}{\partial d_{n}}+\frac{i}{2\theta^{2}\psi_{0n}}
\{\frac{\partial^{2} \psi_{0n}}{\partial \theta^{2}}-\frac{2}{\theta}
\frac{\partial \psi_{0n}}{\partial \theta}\}
\end{equation}
and,
\begin{equation}
\frac{\partial\sigma}{\partial \theta}\approx \frac{5\lambda}{8\theta^{4}}
\end{equation}
Which shows that there seems to exits an apparent divergency for $\Psi^{(2)}$ as
$\theta\rightarrow 0$.
Moreover, from Eqs. (9) and (17), upon factorizing  Van Vleck determinant,
we pick up the $n$ - mode wave function $\psi_{0n}$ as
\begin{equation}
\psi_{0n}=\theta e^{-\frac{1}{2}n a^{2}(\theta)d_{n}^{2}}
\end{equation}
If we now separate the factor ordering dependent part (i.e., that
arising from the Van Vleck determinant) of $\eta_{n}$ in the form
\begin{equation}
\eta_{j}(\theta, d_{j})=\eta_{j1}(\theta, d_{j})+\eta_{2}(\theta)\delta_{nj}
\end{equation}
we finally obtain, replacing Eqs. (22) and (23) in Eq. (20)
\begin{equation}
\frac{\partial}{\partial \theta}[\sigma(\theta)+\eta_{2}(\theta)]=0
\end{equation}
\begin{equation}
\frac{i}{\lambda}\frac{\partial \eta_{n1}}{\partial \theta}=
\frac{1}{2}\frac{\partial^{2}\eta_{n1}}{\partial d_{n}^{2}}-
d_{n} n a^{2}(\theta)\frac{\partial \eta_{n1}}{\partial d_{n}}
\end{equation}
Eq. (24) demostrates that, after the adiabatic approximation,
the divergencies arising in $\sigma$ and
$\eta_{n}$ as we approach $\theta \rightarrow 0$ cancel out each other exactly.
This, on the other hand, is a consequence of the fact that the phase
correcting terms from quantum gravity should not depend on the selection
of the factor ordering \cite{kn:Kiefer}.
Therefore, in $\Psi^{(2)}$, the only physically relevant quantum gravity
correcting phase factor is $\eta_{n1}(\theta,d_{n})$.

The initial state of the Universe is taken on the three-sphere
$a^{2}\rightarrow 1/\lambda$, in this case, we can obtain an exact solution
of Eq. (25) upon making the obvious substitutions
\begin{equation}
\eta_{n1}=e^{i\varepsilon \lambda\theta} y(d_{n}) \mbox{,}
\end{equation}
for some unknown constant $\varepsilon$. Then we  write
\begin{equation}
d_{n}=(\lambda /n)^{1/2}x_{n}     \mbox{,}
\end{equation}
\begin{equation}
k=\varepsilon \lambda /n          \mbox{,}
\end{equation}
using these expresions, Eq. (25) transforms into
\begin{equation}
\frac{d^{2}y}{d x_{n}^{2}}-2x_{n}\frac{d y}{d x_{n}}+2k y=0 \mbox{,}
\end{equation}
which is the Hermite equation. The requirement of normalizability
in $d_{n}$ of the corrected wave function $\psi_{n}=\psi_{0n}e^{i \frac{\eta_n1}{m_{Pl}^{2}}}$
imposes that $\eta_{n1}$ should only be given in terms of polynomials 
in $d_{n}$, i.e., we select the constat $k$ in Eq. (29)
\begin{equation}
k=0,1,2 \mbox{,}
\end{equation}
or
\begin{equation}
\eta_{n1}(\theta, d_{n})=\frac{(2)^{1/2}g_{1}}{2}e^{in\theta/2}
[2(\frac{n}{\lambda})^{1/2}d_{n}\lambda]+
g_{2}e^{i n\theta}[(\frac{4nd_{n}^{2}}{\lambda}-2)\lambda] \mbox{.}
\end{equation}
After Eq. (22), we notice that $g_{1}\neq 0$ in Eq. (31) (i.e., a linear term in the phase   
of the $n$-mode wave fuction of gravitons) 
means that the expected initial number
of gravitons is different from zero and it is given by
\begin{equation}
N=\lim_{a^{2}\rightarrow 1/\lambda}
|\frac{2g_{1}\lambda (n/\lambda)^{1/2} i}{m_{Pl}^{2}(2n a^{2})^{1/2}}|^{2}=
|\frac{g_{1}\lambda}{m_{Pl}^{2}}|^{2}        \mbox{.}
\end{equation}

\section{The spectrum of relic gravitons}

Since the initial number of gravitons  is different
from zero, it will also change the measurable properties of the resulting   
spectrum corresponding to the statistic of gravitons in inflationary models;
here, in order to calculate the expected changes, 
we follow the method of B. Allen\cite{kn:Allen}.

In the first approximation, the graviton spectrum produced 
by an inflationary stage is enterely independent of the
mechanism that produces the inflation. The only inputs which are needed 
to find the graviton spectrum are the classical metric of space-time,
and the initial quantum state of the gravitational perturbations.

For convenience, one may assume that the Universe is approximatelly flat, 
so that the metric takes the form
\begin{equation}
ds^{2}=a^{2}(t)(-dt^{2}+d\sigma^{2})
\end{equation}
In any case, since we will only study the situation 
corresponding to wavelengths which
are shorter than the present-day horizon scale, it could be taken as
a good approximation if the Universe were either spatially open or closed.

The classical spacetime  begins as de Sitter space but then undergoes an 
instantaneous phase transition at, say, $t=t_{1}$, after which it evolves as a 
radiation-dominated model until the time $t=t_{0}$. The scale factor describing
this model is
\begin{eqnarray*}
a(t) =(t/t_{1})a(t_{1}) \,\, \mbox{for}\,\,\, t_{1}<t<t_{0}
\end{eqnarray*}
\begin{equation}
a(t)=(2-t/t_{1})^{-1}a(t_{1}) \,\, \mbox{for}\,\,\, t< t_{1}
\end{equation}

We would also require the solution of Einstein equations, imposing
\begin{equation}
\frac{8\pi G \rho_{0}}{3}=\frac{1}{a(t_{1})^{2}t_{1}^{2}}\equiv \lambda
\end{equation}

Let us now determine the gravitational-particle production 
in this spacetime.

A gravitational perturbation with comoving wave number {\bf k} is represented
by
\begin{equation}
h_{\mu\nu}=a(t)^{2}\epsilon_{\mu\nu}(k)\phi(t)e^{i k x}+ cc 
\end{equation}
where $\epsilon_{\mu\nu}$ is the polarization tensor. The physical
frequency of the wave is $\omega=k/a(t)$. The amplitude $\phi$ also
obeys the perturbed Einstein equations, it  leads to
\begin{equation}
\ddot{\phi}+(2\dot{a}/a)\dot{\phi}+k^{2}\phi=0
\end{equation}
The choice of a solution to the equation for $\phi$ corresponds to the choice
of an initial quantum state for the gravitational field. In de Sitter
stage, the solution representing a de Sitter-invariant vacuum state is
\begin{equation}
\phi_{v}(t)= [a(t_{1})/a(t)]\{1+i(\lambda)^{1/2}(a(t)/k)\}e^{-ik(t-t_{1})}
\end{equation}
On the other hand, if, after Eq. (32), the initial state is given by a small $N$-graviton
coherent state, we must correct the vacuum state upon adding a negative
frequency solution to the above expression,
\begin{equation}
\phi(t)= \phi_{v}(t)+
(i g_{1}\lambda/m_{Pl}^{2} )\phi_{v}^{*}(t)+
O(\lambda/m_{Pl}^{2})^{2}
\end{equation}
The corresponding solution to the wave equation in the radiation stage
is
\begin{equation}
\phi_{r}(t)=(a(t_{1})/a(t))\{\alpha_{r}e^{-ik(t-t_{1})}+\beta_{r}e^{+ik(t-t_{1})}\}
\end{equation}
where $\alpha_{r}$ and $\beta_{r}$ are Bogolubov coefficients.

By matching the modes at $t=t_{1}$ one obtains for the negative frequency coefficient,
representing particle creation
\begin{equation}
\beta_{r}=(-ig_{1}\lambda/m_{Pl}^{2})(1-\frac{i}{kt_{1}})+
\frac{(1+ig_{1}\lambda/m_{Pl}^{2})}{2k^{2}t_{1}^{2}}
\end{equation}
On the other hand, upon taking into account that $k=a(t_{0})\omega$,
$t_{1}=1/a(t_{1})\lambda^{1/2}$ we finally obtain,
for the total number of gravitons at time $t_{0}$,
\begin{equation}
|\beta_{r}|^{2}=\frac{1}{4}[\frac{a(t_{1})}{a(t_{0})}]^{4}
\frac{\lambda^{2}}{\omega^{4}}\epsilon(\omega)
\end{equation}
where,
\begin{equation}
\epsilon(\omega)\equiv\ 1-2(\lambda/m_{Pl}^{2})
[2 Re[g_{1}][\frac{a(t_{1})}{a(t_{0})}] \frac{\omega}{\lambda^{1/2}}+
Im[g_{1}](1-2 [\frac{a(t_{1})}{a(t_{0})}]^{2}\frac{\omega^{2}}{\lambda }]
\end{equation}
Therefore, the spectrum should be  given by the corresponding energy density 
$d\rho_{g}=P(\omega) d\omega$
for a density of states $dN=\omega^{2}d\omega/(2\pi^{2})$, i.e.,
\begin{equation}
d\rho_{g}=P(\omega)d\omega=2 \omega  \frac{\omega^{2}d\omega}{2\pi^{2}}|\beta_{r}|^{2}
\end{equation}
or,
\begin{equation}
P(\omega)=\frac{1}{4\pi^{2}}\frac{\lambda^{2}}{\omega}
[\frac{a(t_{1})}{a(t_{0})}]^{4} \epsilon(\omega).
\end{equation}
Here, $\epsilon(\omega)$ differs from Zel'dovich's spectrum due to the presence
of quantum gravity effects. If we take into account some phenomenological
values of $[\frac{a(t_{1})}{a(t_{0})}]\sim 10^{28}$ (see also \cite{kn:Allen}), 
we conclude that such effects would already  be present for those 
frequencies of the order of $\omega \sim 10^{13}$ Hz.

The result in Eq. (45) may also be expressed in terms of the effective Hubble
parameter during inflation, $H=\lambda^{1/2}$, and the recombination density
$\rho_{R}=\rho_{0} [\frac{a(t_{1})}{a(t_{0})}]^{4}$, then, Eq. (35) taken into
account, we obtain the typical value   
\begin{equation}
\Omega_{graviton}\equiv\frac{\omega}{\rho_{R}}\frac{d\rho_{g}}{d\omega}=
 \frac{2}{3\pi}(\frac{H}{m_{Pl}})^{2}\epsilon(\omega)
\end{equation}
where $\epsilon(\omega)\approx 1+O(\lambda/m_{Pl}^{2})$,
represents a perturbation to the predicted Zel'dovich plateau.

\section{Concluding remarks}

Cosmic Background Radiation is used in order to test
phenomenological models of the early Universe. In these scenarios,
we have just 
shown that, quantum gravity corrections  for the spectrum of gravitons,
would possibly lie on
the  range of frequencies technically accessible in
some future experimental devices  measuring CBR anisotropy\cite{kn:COBRAS}.

Semiclassical gravity, represented by $\Psi^{(2)}$, is, in this framework,
a predictive and testable theory of initial conditions. On the other hand,
we have also seen that
the adiabatic approximation  for the wave function of metric perturbations,
leading to the ground state for the
wave function of gravitons,
can be thought as being technically correct since, in this case, there would not exist
formal divergencies depending on the factor ordering 
for the resulting quantum gravity corrections on the initial three-sphere.  

\section{Acknowledgements}
The author whishes to thank Claus Kiefer 
and Jos\'e Luis S\'anchez-G\'omez  for many useful comments.
This work is supported by the grant EX95-08960718 from the Spanish
Ministry of Education and Culture and the project C.I.C.y T. PB 94-0194.

\end{document}